\documentclass[preprint]{elsarticle}
\journal{Nuclear Physics B}

\usepackage{url}
\usepackage{graphicx}
\usepackage{axodraw2}

\usepackage{latexsym}
\usepackage{amssymb}
\usepackage{amsfonts}

\usepackage{amsmath}
\usepackage[caption=false]{subfig}

\usepackage{breqn}

\newcommand{\ep}{\epsilon}

\newcommand{\dd}{\mbox{d}}

\newcommand{\nn}{\nonumber}

\allowdisplaybreaks[4]
 
\begin{document}
\begin{frontmatter}

\title{
\normalfont
\vskip-1cm{\baselineskip14pt
  \begin{flushleft}
   \end{flushleft}}
  \vskip1.5cm
How to choose master integrals}
 
\author[SRCC,KIT,MCFAM]{A.V.~Smirnov}
\ead{asmirnov80@gmail.com}

\author[SINP,MCFAM]{V.A.~Smirnov\corref{cor1}}
\ead{smirnov@theory.sinp.msu.ru}

\cortext[cor1]{Corresponding author}
 
\address[SRCC]{Research Computing Center, Moscow State University, \\119992 Moscow, Russia}
\address[KIT]{Institut f\"ur Theoretische Teilchenphysik, Karlsruhe Institute of Technology (KIT),
76128 Karlsruhe, Germany}
\address[SINP]{Skobeltsyn Institute of Nuclear Physics of Moscow State University,\\  119992 Moscow, Russia}
\address[MCFAM]{Moscow Center for Fundamental and Applied Mathematics}

\begin{abstract}
The standard procedure when evaluating integrals of a given family of Feynman integrals,
corresponding to some Feynman graph, is to construct an algorithm which provides the
possibility to write any particular integral as a linear combination of so-called 
master integrals. To do this, public ({\tt AIR}, {\tt FIRE}, {\tt REDUZE}, {\tt LiteRed}, {\tt KIRA}) and private codes based on solving integration by parts relations 
are used. However, 
the choice of the master integrals provided by these codes is not always optimal.
We present an algorithm to improve a given basis of the master integrals, as well as its computer
    implementation; see also a competitive variant \cite{Usovitsch:2020jrk}.
\end{abstract}
\end{frontmatter}
\newpage
%

\section{Introduction}

After integration by parts (IBP) reduction was invented~\cite{Chetyrkin:1981qh}
it became possible to decompose the problem of evaluating Feynman integrals
into two parts: a reduction to so-called master integrals (MIs) and the evaluation of these
MIs. In the eighties and nineties, the first part of this procedure was solved `by hand'
but then computer codes which perform an IBP reduction appeared.
At the moment, there are at least five public 
codes ({\tt AIR}, {\tt FIRE}, {\tt REDUZE}, {\tt LiteRed}, {\tt KIRA})
~\cite{Anastasiou:2004vj,Smirnov:2014hma,Smirnov:2019qkx,vonManteuffel:2012np,Lee:2012cn,Lee:2013mka,Maierhoefer:2017hyi,Maierhofer:2018gpa}, and a number of
private codes\footnote{Let us observe that, for concrete families of Feynman integrals,
specially constructed IBP-reduction programs can be much more powerful than the above mentioned
general programs. Here remarkable examples are two 
public codes {\tt Mincer} \cite{Larin:1991fz,Gorishnii:1989gt} and {\tt Forcer} \cite{Ruijl:2017cxj}
successfully applied for the reduction of three- and four-loop massless propagator diagrams, respectively.}. By definition, MIs are integrals that appear on the right-hand sides
of solutions of IBP relations, so that they form a basis in the linear space of integrals of a given family associated
with an $h$-loop graph, 
\begin{eqnarray}
G_{i_1,\ldots,i_L}
=
\int\ldots\int \prod_{l=1}^L \frac{1}{(m_l^2-p_l^2)^{i_l}}
\dd^d k_1\ldots \dd^d k_h  \,.
\label{fam}
\end{eqnarray}
Here $d=4-2\ep$ is the dimensional regularization parameter, $k_1,\ldots,k_h$ are loop momenta,  
and momenta of the lines $p_l$ are expressed in terms of linear combinations of the loop momenta $k_i$ and
external momenta $q_j$. Integrals of a given family are, in particular, functions of indices $i_l$
(powers of the propagators) which can be considered as integer variables.
 
The title of the paper might look strange because a set of the MIs is produced automatically after a code to solve IBP relations is applied, so that there is no choice at this point. However, experience tells us that, especially in sufficiently complicated situations, the basis provided by such a code, can be bad because the denominators of the coefficients of MIs
in IBP-reductions of input integrals can be quite cumbersome. Of course, coefficients in the decomposition of
a given input integral over MIs are always rational functions of everything, i.e. of $d$ and kinematical invariants, because solving IBP relations reduces to solving sparse linear systems of equations with the help
of a variant of the Gaussian elimination. With big denominators, the reduction to the MIs can be rather complicated and, in some cases, even unfeasible, i.e. requiring too much time or/and operative memory. 

It looks natural to expect that
the denominators in IBP-reductions are connected with singularities of Feynman integrals 
whose position follows from an analysis of convergence properties of Feynman integrals represented as parametric integrals over Feynman parameters.
This analysis can be performed, in some situations, with classical sector decompositions by Hepp and
Speer used to prove theorems on renormalization~\cite{Hepp:1966eg,Speer:1975dc}, or, 
in more general situations, with modern recursive sector
decompositions~\cite{Binoth:2000ps,Bogner:2007cr,Smirnov:2008py}. 
From this analysis, it follows that the singular factors are either functions of kinematical invariants and 
masses (independent of $d$) described by Landau equations, or linear functions of $d$ 
(independent of other variables). 

Such standard singular factors in denominators of IBP-reductions are unavoidable but we could try to eliminate all more complicated factors using a transition to an appropriate basis of the MIs in which denominators on the right-hand side of IBP reduction relations will be {\em good},
i.e. decomposed as products of polynomials of kinematical invariants and masses, independent of $d$, and
linear terms of the form $a d+b$ with rational numbers $a$ and $b$. 
Let us also call a denominator {\em bad} if it is not good.
We will also call a basis {\em good} if the denominators in IBP reductions into that basis are good.

In fact, the possibility of finding a good basis follows from\footnote{We are grateful to Erik Panzer who turned our attention to this theorem after the archive version of this paper appeared.} 
Theorem~0.6 of Ref.~\cite{BSMF_1992__120_3_371_0} by Sabbah about the solution of
a system of difference equations of several variables. 
The main ingredients of the corresponding formalism
are shift operators with respect to these variables and operators of multiplication by these 
variables. Similarly, the shift operators for the indices of Feynman integrals and the corresponding 
multiplication operators are standard ingredients of IBP relations.\footnote{A description of Feynman integrals in terms of a vector space of rational functions can be found, e.g., in
Ref.~\cite{Bitoun:2017nre}.}
We apply this theorem to a family of Feynman integrals and the corresponding IBP relations are
difference equations. As the variables we have the indices (which we consider integer) and dimension, 
$\{i_1,\ldots,i_L,d\}$. The evolution of Feynman integrals under the action of the shift operators 
is described by multiplication by matrices (representing the shift operators) 
composed of rational functions of our variables.
We can apply this theorem because the basis of MIs is finite dimensional~\cite{Smirnov:2010hn}.
Now, the theorem states that there is always a basis such that the matrices representing the shift operators
have denominators which are products of linear functions of indices and dimension. 
We can obtain a given integral of the given family
by the action of a finite number of shift operators on the elements of the basis of the master integrals,
so that this integral can be obtained from the basis of the master integrals by the action
of a matrix composed of rational functions.
Therefore, according to the the Sabbah's theorem, there should be a basis of the master integrals such that there are no bad denominators in results of IBP reductions.

Guided by the existence of a good basis provided by the Sabbah's theorem, 
we are now going to explain how one can {\em practically} improve a given basis of the MIs if it is not 
good\footnote{In our experience, we already improved bases of MIs in many calculations
without developing a code for this and we believe that other people also did this.
Here is one more example from the literature~\cite{Melnikov:2016qoc}.}. 
In the next section, we describe an algorithm to improve a given basis of MIs. 
In Section~3, we discuss possible origins of bad denominators.
In Section~4, we present a code based on our algorithm and, in Section~5, we discuss some other ways of improving a given basis of MIs.
In the Appendix, we present an example which demonstrates how our code works.

\section{The algorithm}  
  
Suppose we have a basis $f_i(x,d), i=1,...,N$, of MIs obtained with some IBP reduction code,
 For simplicity of presentation, we describe the case of two scales, where
$x$ is their ratio, for example, $x=q^2/m^2$.
Let us check whether it is a good or bad basis and if it is bad let us try to 
improve it. Let us run an IBP reduction code on a set of sample 
integrals taken from all the sectors with non-zero numbers of the MIs.
In our calculations, we prefer to choose integrals with indices 0,1 and 2:
we include in the sample list corner integrals of these sectors, i.e. without indices equal to two,
then integrals with one index equal to two, then integrals with two indices index equal to two.
(In complicated situations, sample integrals with three indices equal to two or even higher might 
be also needed.) We prefer sample integrals without negative indices because, according to our experience, 
the choice of MIs with negative indices has more chances to lead to an appearance of bad denominators.
Moreover, symmetries of Feynman integrals are more visible for integrals without negative indices.
However, the following algorithm and its implementation work for any set of sample integrals.

Anyway, we start with an IBP reduction of a set of the sample integrals and know their reduction
which can be written in the form of a list of substitutions:
\begin{eqnarray}
f(x,d) \to \sum_{i=1}^N c_i(x,d) f_i(x,d) \;,
\end{eqnarray}
where the coefficients $c_i$ are rational functions of $x$ and $d$.
 
Let us call by the {\em level} the number of positive indices of an integral.
Let us analyze reductions of the sample integrals starting from sectors of
the minimal level. Suppose that we are at the lowest level where bad denominators appear.
We now determine which sectors are responsible for the generation of these bad 
denominators by analyzing at which of the MIs of the given level the bad denominators appear.
We now consider these sectors one by one.
 
For a given sector $\sigma$, let $g_j, j=1,2,...,$ be the set of the corresponding sample
integrals.
Their reduction has the form
\begin{eqnarray}
g_j=\sum_{i=1}^{|\sigma|} c_{j,i} f_i + \ldots, 
\end{eqnarray}
where $f_i, i=1,2,...,|\sigma|$ are MIs of the 
given sector and dots stand for the contribution of lower sectors.
For a given $g_j$, analyze numerators of those coefficients $c_{j,i}$ which
involve the current bad denominator. Let us consider a numerator bad or good 
using the same definition as formulated above for the denominators.

{\em (a) A simple situation}. Suppose that for some $i$, the numerator is good. Then replace the MI $f_i$ by the new MI $g_j$.
After using an explicit relation between $f_i$ and $g_j$ which is found by solving a linear equation, 
find the mapping which expresses $f_i$ in terms of $g_i$ and the other MIs.
Check that after this change, the current bad denominator disappears.
When $f_i$ is written down in terms of $g_i$ and other MIs, the bad denominator
goes to the numerator and cancels bad denominators also in other places, while
the numerator in $c_{j,i}$ goes to the denominator but it is harmless because
it is good. 
After a transition to the current new MIs, the code checks that the bad denominators
under consideration disappear in the IBP reduction of {\em all} the sample integrals of the
given sector.

{\em (b) A more complicated situation}. Suppose now that for all $i$, the corresponding numerators are bad.
Choose $i$ such that the length (defined as the number of terms in the expanded expression) of the numerator is minimal and make the corresponding replacement.
Therefore, the resulting bad denominators become better, but they are not yet good. Repeat this procedure until all denominators are good.

Now perform this procedure also for other sectors of the given level, then proceed to
higher levels eventually reaching the top sector.  
As a result we obtain a list of desirable MIs. Within FIRE, this list is encoded via the option {\tt preferred}
in subsequent reductions.
    
To speed up the analysis of the bad denominators, one can fix either $d$ or other variables.
In particular, in situations with many kinematic invariants, one can get rid of non-linear denominators in $d$
fixing all the other parameters and thereby make the sample reduction much faster.
This procedure is systematically described in an alternative version~\cite{Usovitsch:2020jrk} of getting rid of bad denominators.

\section{Where do the bad denominators appear from?}

In order to use the code efficiently and not to expect it to do things that it is not designed to do, it is important to understand how the bad denominators appear. To our understanding, 
bad denominators can appear because either

\begin{enumerate}
    \item the current choice of MIs is not the optimal choice, or,
    \item the current set of MIs is not minimal so that there is a hidden relation between them.
\end{enumerate}

It is important to understand which of those variants (or both) is the case to improve properly the current
basis of MIs. 

But what actually is this variant 2? If MIs are irreducible how can one have a relation between them? The answer lies in the implementation of reduction programs. There can be a relation between "MIs" produced by a reduction program that the reduction program cannot reveal, and there are reasons for this, one of which is that it might be that not all relations between Feynman integrals follow from IBP relations. In fact, it is an open question whether they follow or not. However modern reduction programs normally try to use symmetries in addition to the IBP relations. For example, FIRE can use internal sector symmetries from LiteRed (depending on the $\#pos\_pref$ option), and so we normally do not miss relations for MIs in a single sector.

Still, there is a reason why extra relations can be missed due to the way in which reduction programs are implemented. (The statement is valid for FIRE, but we expect it is also to be valid for other reduction programs.) The programs work sector by sector, so if during reduction a relation is reduced completely out of a sector, relating only integrals of lower sectors, reduction programs tend to drop relations of this sort at this point. Therefore we locate a possible source of relations between MIs of lower sectors that reduction programs can ignore.

This leads to the following conclusion: the variants can be distinguished one from the other. In case there is a relation $r$ between MIs of level $l$, it means that there should be a relation (IBP or symmetry) of a level higher than $l$, that could be reduced and lead to $r$. It also means that the analysis of bad denominators at level $l$ won't reveal such a relation. The bad denominators of type 2 are revealed only when one takes a reduction relation for an integral of a level higher than $l$, and the bad coefficients are those at integrals of level $l$.

On the other hand, the bad denominators of type~1 appear inside a level, when considering coefficients of integrals of the same level on the left-hand side and right-hand side of the formula. Still due to the way we order Feynman integrals (trying to reduce to lower sectors), there will be bad denominators at lower levels as well.

It is important to note, that the code described here only aims at a good basis choice 
and tries to get rid of bad denominators of the first type. In case there are extra relations between MIs, one needs another way to decrease their number, and this will be discussed in Section~5, however even in this case the code can improve the basis.

However this consideration has another important consequence that might be useful for the application of the code: while searching for bad denominators of type 1, one can consider only coefficients expressing integrals of a given level by MIs of the same level. Everything that is below can be dropped for the purpose of finding a good basis.

\section{The code} 
 
The above algorithm is implemented in Mathematica as a part of FIRE, starting from the public release 6.4.1 (with more options in 6.4.2), however all functions related to this algorithm are placed in a separate Mathematica file mm/ImproveMasters.m, and moreover it can be used not only together with FIRE, but also with other reduction programs.

Let us explain the format used by the algorithm. First, there is a Feynman integral, which is defined by a problem number $pn$ (a positive integer) and a set of indices. Like in other parts of FIRE, we use the following form for a Feynman integral:
\begin{equation}
    G[pn, \{i_1, i_2, \ldots, i_n\}]
\end{equation}

Then let us define a `relation', i.e. a representation of one integral as a linear combination of other integrals. Of course this could be simply a Mathematica rule with a linear combination on the right-hand side, but for optimization reasons we prefer to store a relation in a structured format, where the right-hand side of the rule (Mathematica \textit{Rule}) comes as a list of pairs containing an integral and a coefficient each:

\begin{eqnarray}
    G[pn, \{i_1, i_2, \ldots, i_n\}] \rightarrow \nonumber\\ \{\{G[pn, \{j_{1,1},\ldots,j_{1,n}\}], c_1\},\ldots,\{G[pn, \{j_{m,1},\ldots,j_{m,n}\}], c_m\}\}
\end{eqnarray}

The problem number should be always the same, the integrals on the right-hand side should not be repeated. This format is much more convenient for algorithmic reasons because one does not need to separate coefficients from the right-hand sides all the time. It is also not difficult to convert between the traditional format with a sum and the structured format. We provide a function $RelationSum2List$ that converts a rule with a sum to a rule with a list. The inverse conversion is even less complex and can be obtained with $Rule[\#\#[[1]], Plus @@ Times @@@ \#\#[[2]]]\&$, but we also for convenience we provide the $RelationList2Sum$ function.

The input for the main algorithm is a Mathematica list of relations. This format can be obtained in FIRE with the 
\begin{equation}
Tables2Rules[filename, Identity, False]
\end{equation}    
command. Here $Identity$ stands for no function application to coefficients, this will be done by the algorithm later anyway. The last parameter $False$ stands for $JoinTerms = False$ meaning that we are not going to convert the expressions from the list format to the sum format.

The main function provided by the algorithm is
\begin{equation}
ImproveMasters[relations, level]
\end{equation}

or, starting from version 6.4.2,

\begin{equation}
ImproveMasters[relations, level,length]
\end{equation}

Here $level$ stands for the level of integrals (number of positive indices) in which the code will work,
and $length$ is the minimal length of a polynomial independent of $d$ starting from which 
it is considered bad. For example, $length=10$ can be a reasonable choice.

The intermediate output of the code is self-explanatory; it prints the bad denominator factors found, the sectors in which they are found, lists of MIs involved and the replacements of MIs it makes in order to get rid of bad denominators. The output is a pair containing a new set of relations and the good basis of MIs. The list of MIs in the output contains only MIs in sectors where a change was required. This set can be used as the set of preferred MIs in FIRE in subsequent IBP reductions or in a similar way in other reduction programs.

To find bad factors FIRE uses \textit{Together} to simplify the fractions, then uses the \textit{Denominator} function to get the denominators and then calls the provided $FindBadFactorsInCoefficient$ function that first uses \textit{FactorList} and then analyses the factors. The condition for the factor to be "bad" is a set either depending both on $d$ and other variables or containing a non-linear dependence on $d$.

As explained in the previous section, the search for a good basis can be performed purely inside a given level, dropping everything that is below. Therefore if one is not immediately interested in new relations but is searching for the list of preferred integrals only, it is safe to restrict the search to a given level, and that can improve performance of the code greatly. To do that one can use the function $LevelPart[relations,level]$ that keeps only the current level part. For example, one can call

\begin{equation}
    ImproveMasters[LevelPart[relations,level],level][[2]]
\end{equation}
to get the list of preferred integrals for the current level. Note that the $LevelPart$ function not only picks the relations for integrals of a given level (which could be done, for example, with 

\noindent $Select[relations, (IntegralLevel[First[\#\#]] == level)\&]$) but also leaves only integrals of the desired level on the right-hand side.

The code comes with a number of auxiliary functions. The 
\begin{equation}
BadRelationParts[expr\_, level\_:0, onlyCurrent\_:False]
\end{equation}
function picks out only parts of the original rules where coefficients contain a bad factor in the denominator. If $level$ is non-zero, then it considers only integrals of the specified level on the left-hand sides, and if also $onlyCurrent$ is set to $True$, then it also filters the right-hand sides to have only integrals of the same level. This might be useful since only coefficients at the current level are related to the search of a good basis. In case the function $ImproveBasis$ succeeds, the result of $BadRelationParts[expr, level, True]$ on the first part of its return value be an empty set. 

If one is only interested in displaying bad factors, ignoring the integrals with which they appear, one can use the 
\begin{equation}
FindBadFactorsInRules[expr\_, level\_:0, onlyCurrent\_:False]
\end{equation}
function with the same parameters. Similarly, if the function $ImproveBasis$ succeeds, the result of $FindBadFactorsInRules[expr, level, True]$ on the first part of its return value should be an empty set.

The purpose of the code is in searching for a good basis, and this is related only to coefficients of a given level. As explained above there can be extra relations between MIs that usually lead to bad coefficients of MIs of a lower level than the integral on the left-hand side. Since the code $ImproveBasis$ cannot help with getting rid of them, alternative methods should be used that are discussed briefly below. However if one has such an extra relation, it can also be applied with the structured rules format. This is done with the $SubstituteRuleIntoRules[rules\_, rule\_]$ function.

In case there are no more bad denominators remaining, both functions $BadRelationParts[expr]$ and $FindBadFactorsInRules[expr]$ return empty sets.

\section{Discussion and conclusion}

We have explained how to get rid of bad denominators by improving a given basis of the MIs.
Suppose now that some bad denominators survive after using our code. 
As mentioned above, another source (in addition to the choice of an improper basis of the MIs)
of bad denominators can be hidden relations between a current set of the MIs.
When presenting the release of {\tt FIRE4} \cite{Smirnov:2013dia} we suggested a way to find some 
of such relations.
It is based on symmetries of Feynman integrals of a given family. One finds relations between sample integrals
(we prefer to consider integrals with indices equal to 0, 1 and 2, with the number of indices $=2$ equal to one, two, three and, in some complicated cases, even higher, however the code can work with any sample choice). Within {\tt FIRE}, this can be done with the help of the command {\tt FindRules} but there are also other ways to find such symmetry relations.
Then one performs an IBP reduction of these sample integrals and checks whether symmetry relations
simply yield identities or produce new relations between current MIs.
However, there exist hidden relations which cannot be revealed by this procedure.
A first example of such nontrivial relations and their description is presented in 
Ref.~\cite{Georgoudis:2020p5l}.

In fact, one can check whether the number of the MIs in a given sector is minimal using the code 
{\tt Mint} \cite{Lee:2013hzt} based on algebraic geometry. If the code gives a number which is less than the number of current MIs in the given sector then it is quite reasonable to look for a hidden relation.
However, additional relations obtained with the help of symmetries usually provide relations in
partially overlapping sectors while {\tt Mint} provides information about a fixed sector.

We also know examples where running an IBP reduction with {\tt KIRA} and {\tt FIRE} and equating the corresponding results provides a missing relation.

Getting rid of bad denominators is important not only because it improves reduction performance with respect to runtimes and memory usage.
In particular, when applying approaches based on modular arithmetic 
({\tt Finred} \cite{vonManteuffel:2014ixa}, {\tt KIRA}~\cite{Usovitsch:2020jrk} and {\tt FIRE} \cite{Smirnov:2019qkx}), 
it is necessary, first, to reveal the form of possible denominators.
Moreover, within the method of differential equations, it is important to get rid of
denominators which are spurious and can be eliminated by a basis change.

Let us emphasize that our algorithm can be applied not only with {\tt FIRE} but also with other reduction programs.
On the other hand, one more tool for improving a given basis of MIs is described in
a `parallel' paper~\cite{Usovitsch:2020jrk} by an author of {\tt KIRA}.

Let us point out that bad denominators appear only at a sufficient level of complexity, so that
one does not need our code in simple cases. 
However, in complicated cases, the code can essentially improve the situation with the IBP reduction.
Our approach is pragmatical: we do not prove that, under some conditions, it should work.
We accept that, in some complicated situations, the code can 
meet difficulties so that it will be necessary to develop it further or just to include
more sample integrals into the game, in addition to the described default choice, i.e.
various sample integrals with the indices 0,1 and up to two indices equal to 2.
However, we have already applied our heuristic code in several projects and believe that the readers will also successfully apply it in practice.

\appendix

\section{An example}

Let us see how bad denominators can be eliminated by our code for the family of integrals associated with the 
three-loop vertex graph shown in Fig.~\ref{fig_example}.
\begin{center}
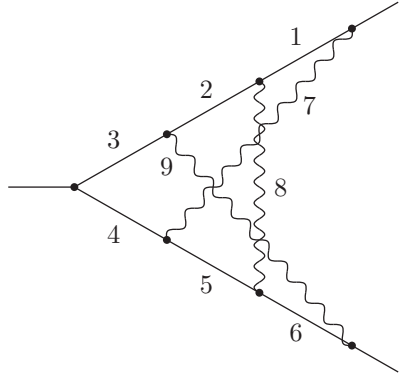
\begin{figure}[ht] 
\begin{picture}(120,140)(-100,-70)
\Line(35,20)(70,40)
\Line(70,-40)(35,-20)
\Line(35,-20)(0,0)
\Line(0,0)(35,20)
\Line(-25,0)(0,0)
\Line(70,40)(122.5,70)
\Line(70,-40)(122.5,-70)
\Photon(35,20)(105,-60){2}{10}
\Photon(70,40)(70,-40){2}{10}
\Photon(35,-20)(105,60){2}{10}
\Vertex(0,0){1.5}
\Vertex(35,20){1.5}
\Vertex(35,-20){1.5}
\Vertex(70,40){1.5}
\Vertex(70,-40){1.5}
\Vertex(105,60){1.5}
\Vertex(105,-60){1.5}
\Text(84,57)[]{$1$}
\Text(84,-55)[]{$6$}
\Text(15,18)[]{$3$}
\Text(15,-18)[]{$4$}
\Text(50,37)[]{$2$}
\Text(50,-37)[]{$5$}
\Text(78,0)[]{$8$}
\Text(35,8)[]{$9$}
\Text(89,31)[]{$7$}
\end{picture}
\caption{\label{fig_example}
A three-loop vertex graph. Solid lines are with the mass $m$, wavy lines are massless.}
\end{figure}
\end{center}
In accordance with Eq.~(\ref{fam}), the squares of momenta $p_l$ in the propagators are
\begin{eqnarray}
\left\{-(q_1 + k_1)^2 + m^2, -(q_1 + k_1 + k_2)^2 + m^2, -(q_1 + k_1 + k_2 + k_3)^2 + m^2,
\right. \nn \\
-(q_2 + k_1 + k_2 + k_3)^2 +  m^2, -(q_2 + k_2 + k_3)^2 + m^2, -(q_2 + k_3)^2 + m^2,
\nn \\ \left.
-k_1^2, -k_2^2, -k_3^2, -(k_1 - k_2)^2, -(k_1 - k_3)^2, -(k_2 - k_3)^2\right\}\;,
\nn 
\end{eqnarray}
where $k_i$ are loop momenta and $q_i$ are external momenta with
$q_1^2 = m^2, q_2^2 = m^2, (q_1-q_2)^2 = s$.
The first nine indices can be positive while the last three indices are always non-positive and stand only for numerators.

As a list of sample integrals we choose the list described in Section~2 without negative indices and
the number of indices equal to $2$ up to two.
Looking for hidden relations between primary MIs,
according to the procedure mentioned in the beginning of Section~5 and based on symmetries,
we find the following relations between MIs in partially overlapping sectors:
\begin{eqnarray}
G_{0,1,0,0,1,2,1,0,0,0,0,0}\to \frac{d-2}{8 m^2}G_{0,0,0,1,0,1,1,1,0,0,0,0}
-\frac{2 d-5}{4 m^2}G_{0,1,0,0,1,1,1,0,0,0,0,0}\;.
\nn
\end{eqnarray}

Tables obtained with {\tt FIRE} for sample integrals can be downloaded from 
\url{http://theory.sinp.msu.ru/~smirnov/imi}.
There is also the set of the MIs which we obtain after improving a primary basis.
Running the code described in Section~4 at levels~3 and~4 shows no bad denominators.
At level~5, there are several bad denominators.
To get rid of the bad denominator
\begin{eqnarray}
448 m^4-240 d m^4+32 d^2 m^4-580 m^2 s+320 d m^2 s
\nn \\
-44 d^2 m^2 s+268 s^2-150 d s^2+21 d^2 s^2
\nn
\end{eqnarray}
we have to make a change in the sector with the following MIs:
\begin{eqnarray}
\{G_{0,1,0,1,1,0,0,1,1,0,0,0},G_{0,1,0,1,1,0,0,1,2,0,0,0},G_{0,1,0,1,1,0,0,2,1,0,0,0},
\nn\\
G_{0,1,0,1,1,0,0,2,2,0,0,0},G_{0,1,0,1,2,0,0,1,1,0,0,0},G_{0,1,0,1,2,0,0,1,2,0,0,0},G_{0,1,0,2,1,0,0,1,1,0,0,0}\}.
\nn
\end{eqnarray}
This can be achieved by choosing $G_{0,2,0,1,2,0,0,1,1,0,0,0}$ instead of $G_{0,1,0,1,1,0,0,2,2,0,0,0}$.
 
To get rid of the bad denominator $12 m^2-4 d m^2-16 s+5 d s$
we have to make a change in the sector with the following MIs:
\[\{G_{0,0,1,0,1,0,1,1,1,0,0,0},G_{0,0,1,0,1,0,1,1,2,0,0,0},G_{0,0,1,0,1,0,2,1,1,0,0,0},G_{0,0,1,0,2,0,1,1,1,0,0,0}\}.\]
This can be achieved by choosing $G_{0,0,2,0,1,0,1,1,1,0,0,0}$ instead of $G_{0,0,1,0,1,0,1,1,1,0,0,0}$.

To get rid of the bad denominator $16 m^2-4 d m^2-10 s+3 d s$
we have to make a change in the sector with
\[\{G_{0,1,0,1,1,1,0,0,1,0,0,0},G_{0,1,0,1,1,1,0,0,2,0,0,0}\}.\]
This can be achieved by choosing $G_{0,2,0,1,1,1,0,0,1,0,0,0}$ instead of $G_{0,1,0,1,1,1,0,0,2,0,0,0}$.
 
To get rid of the bad denominator $28 m^2-8 d m^2-4 s+d s$
we have to make a change in the sector with
\[\{ G_{0,1,1,1,1,0,0,1,0,0,0,0},G_{0,1,1,1,1,0,0,2,0,0,0,0},G_{0,1,1,1,2,0,0,1,0,0,0,0}\}.\]
This can be achieved by choosing $ G_{0,2,2,1,1,0,0,1,0,0,0,0}$ instead of $ G_{0,1,1,1,1,0,0,1,0,0,0,0}$. 
 
To get rid of the bad denominator $4 m^2+4 s-d s $
we have to make a change in the sector with
\[\{ G_{0,1,0,0,1,1,1,0,2,0,0,0},G_{0,1,0,0,1,1,2,0,1,0,0,0}\}.\]
This can be achieved by choosing $G_{0,2,0,0,1,1,1,0,1,0,0,0}$ instead of $G_{0,1,0,0,1,1,1,0,2,0,0,0}$.  
 
At level 6, there is a very bad denominator
\begin{eqnarray}
1166901120 m^6-2228472576 d m^6+1889043552 d^2 m^6-934622944 d^3 m^6
\nn\\
+298051104 d^4 m^6-63674944 d^5 m^6+9133728 d^6 m^6-850048 d^7 m^6
\nn\\
+46656 d^8 m^6-1152 d^9 m^6-1077693120 m^4 s+2033788512 d m^4 s
\nn\\
-1698824192 d^2 m^4 s+825710264 d^3 m^4 s-257891900 d^4 m^4 s+53810944 d^5 m^4 s
\nn\\
-7523356 d^6 m^4 s+681848 d^7 m^4 s-36472 d^8 m^4 s+880 d^9 m^4 s
\nn\\
+323477760 m^2 s^2-602689792 d m^2 s^2+493601656 d^2 m^2 s^2-233171456 d^3 m^2 s^2
\nn\\
+69995306 d^4 m^2 s^2-13842696 d^5 m^2 s^2+1802914 d^6 m^2 s^2-149054 d^7 m^2 s^2
\nn\\
+7094 d^8 m^2 s^2-148 d^9 m^2 s^2-31819200 s^3+58544600 d s^3
\nn\\
-46779452 d^2 s^3+21211150 d^3 s^3-5972893 d^4 s^3+1070399 d^5 s^3-
\nn\\
119334 d^6 s^3+7576 d^7 s^3-210 d^8 s^3 
 \nn
 \end{eqnarray}
which is generated by the sector with ten MIs
\begin{eqnarray}
\{ 
G_{0,1,1,0,1,1,1,1,0,0,0,0},G_{0,1,1,0,1,1,1,2,0,0,0,0},G_{0,1,1,0,1,1,2,1,0,0,0,0},G_{0,1,1,0,1,1,2,2,0,0,0,0},
 \nn\\
G_{0,1,1,0,1,2,1,1,0,0,0,0},G_{0,1,1,0,1,2,1,2,0,0,0,0},G_{0,1,1,0,1,2,2,1,0,0,0,0},G_{0,1,1,0,2,1,1,1,0,0,0,0},
\nn\\  
G_{0,1,2,0,1,1,1,1,0,0,0,0},G_{0,2,1,0,1,1,1,1,0,0,0,0}
\}.
\nn
\end{eqnarray}
The code does not find a replacement that immediately removes this denominator. Then the code looks
for variants of reducing the length of this bad denominator.
The best variant corresponds to a reduction of the length from 35 to 16 and 
is achieved by choosing $G_{0,1,2,0,2,1,1,1,0,0,0,0} $ instead of $ G_{0,1,1,0,1,2,1,2,0,0,0,0}$.

Then the code takes care of the bad denominator 
\begin{eqnarray}
 52992 m^4-47136 d m^4+16784 d^2 m^4-3136 d^3 m^4+328 d^4 m^4-16 d^5 m^4
\nn\\ 
 -30480 m^2 s+24512 d m^2 s-7308 d^2 m^2 s+1036 d^3 m^2 s-82 d^4 m^2 s
 \nn\\
 +4 d^5 m^2 s+4200 s^2-3140 d s^2+774 d^2 s^2-63 d^3 s^2
\nn
\end{eqnarray}
in the previous sector where one of the MIs was replaced.
 This denominator is eliminated by choosing $ G_{0,2,1,0,1,1,2,1,0,0,0,0}$ instead of $G_{0,1,1,0,1,1,2,2,0,0,0,0} $. 
 
To get rid of the bad denominator $3480 m^4-1860 d m^4+244 d^2 m^4-2872 m^2 s+1528 d m^2 s-200 d^2 m^2 s+390 s^2-207 d s^2+27 d^2 s^2 $
we have to make a change in the sector with
\begin{eqnarray}\{ G_{0,1,0,1,1,0,1,1,1,0,0,0},G_{0,1,0,1,1,0,1,1,2,0,0,0},G_{0,1,0,1,1,0,1,2,1,0,0,0},G_{0,1,0,1,1,0,2,1,1,0,0,0},
\nn \\
G_{0,1,0,1,2,0,1,1,1,0,0,0},G_{0,1,0,2,1,0,1,1,1,0,0,0},G_{0,2,0,1,1,0,1,1,1,0,0,0}\}.
\nn
\end{eqnarray}
This can be achieved by choosing $G_{0,2,0,1,2,0,1,1,1,0,0,0} $ instead of $G_{0,1,0,1,1,0,1,1,2,0,0,0}$.

To get rid of the bad denominator $ 16 m^2-4 d m^2-10 s+3 d s$
we have to make a change in the two sectors with
\begin{eqnarray}
\{  
G_{1,0,0,1,1,1,0,1,1,0,0,0},G_{1,0,0,1,1,1,0,1,2,0,0,0},G_{1,0,0,1,1,1,0,2,2,0,0,0},G_{1,0,0,1,2,1,0,1,1,0,0,0},
\nn \\
G_{1,1,0,0,1,1,1,0,1,0,0,0},G_{1,1,0,0,1,1,1,0,2,0,0,0},G_{1,1,0,0,1,1,2,0,2,0,0,0},G_{1,1,0,0,1,2,1,0,1,0,0,0} 
 \}.
\nn
\end{eqnarray}
This can be achieved by choosing $ G_{2,0,0,2,1,1,0,1,1,0,0,0}$ instead of $ G_{1,0,0,1,1,1,0,2,2,0,0,0}$ 
and $G_{2,2,0,0,1,1,1,0,1,0,0,0} $ instead of $ G_{1,1,0,0,1,1,2,0,2,0,0,0}$.

To get rid of the bad denominator $22 m^2-6 d m^2-15 s+3 d s$
we have to make changes in the two sectors with
\begin{eqnarray}
\{  
G_{0,1,0,0,1,1,1,1,1,0,0,0},G_{0,1,0,0,1,1,1,2,1,0,0,0},G_{1,0,0,1,0,1,1,1,1,0,0,0},G_{1,0,0,1,0,1,2,1,1,0,0,0}\}.
\nn
\end{eqnarray}
This can be achieved by choosing $ G_{0,1,0,0,2,1,1,1,1,0,0,0}$ instead of $G_{0,1,0,0,1,1,1,1,1,0,0,0}$ 
and $G_{2,0,0,1,0,1,1,1,1,0,0,0}$ instead of $G_{1,0,0,1,0,1,1,1,1,0,0,0}$.

To get rid of the bad denominator $28 m^2-18 s+3 d s$
we have to make changes in the two sectors with
\begin{eqnarray}
\{  
G_{0,1,1,1,1,0,0,1,1,0,0,0},G_{0,1,1,1,1,0,0,1,2,0,0,0},G_{1,0,1,1,0,1,0,1,1,0,0,0},
G_{1,0,1,1,0,1,0,1,2,0,0,0}\}.
\nn
\end{eqnarray}
This can be achieved by choosing $G_{0,2,1,1,1,0,0,1,1,0,0,0} $ instead of $G_{0,1,1,1,1,0,0,1,2,0,0,0} $ 
and $ G_{1,0,1,1,0,2,0,1,1,0,0,0}$ instead of $G_{1,0,1,1,0,1,0,1,2,0,0,0} $.

To get rid of the bad denominator $8 m^2-2 d m^2-7 s+2 d s $
we have to make a change in the sector with
\[\{ G_{0,1,0,1,0,1,1,1,1,0,0,0},G_{0,1,0,1,0,1,1,1,2,0,0,0}\}.\]
This can be achieved by choosing $G_{0,2,0,1,0,1,1,1,1,0,0,0} $ instead of $G_{0,1,0,1,0,1,1,1,2,0,0,0}$. 
 
At level~7, there is a bad denominator $4480 m^6-4192 d m^6+1280 d^2 m^6-128 d^3 m^6-1040 m^4 s+1208 d m^4 s-456 d^2 m^4 s+56 d^3 m^4 s+120 m^2 s^2-152 d m^2 s^2+66 d^2 m^2 s^2-10 d^3 m^2 s^2+4 d s^3-4 d^2 s^3+d^3 s^3$.
We make a change in the sector with
\[\{G_{0,1,1,0,1,1,1,1,1,0,0,0},G_{0,1,1,0,1,1,1,1,2,0,0,0}\}.\]
This can be achieved by choosing $G_{0,1,1,0,2,1,1,1,1,0,0,0} $ instead of $G_{0,1,1,0,1,1,1,1,2,0,0,0}$.
 
To get rid of the bad denominator $360 m^4-152 d m^4+16 d^2 m^4+16 s^2-8 d s^2+d^2 s^2$
we have to make a change in the sector with
\begin{eqnarray}\{G_{1,0,1,1,0,1,1,1,1,0,0,0},G_{1,0,1,1,0,1,1,1,2,0,0,0},G_{1,0,1,1,0,1,1,2,1,0,0,0},
\nn\\
G_{1,0,1,1,0,1,1,2,2,0,0,0},
G_{1,0,1,1,0,2,1,1,1,0,0,0}\}.
\nn
\end{eqnarray}
This can be achieved by choosing $G_{2,0,1,1,0,1,2,1,1,0,0,0} $ instead of $G_{1,0,1,1,0,1,1,2,2,0,0,0} $.  
 
To get rid of the bad denominator $360 m^4-152 d m^4+16 d^2 m^4+16 s^2-8 d s^2+d^2 s^2 $
we have to make a change in the sector with
\begin{eqnarray}
\{ G_{1,0,1,1,0,1,1,1,1,0,0,0},G_{1,0,1,1,0,1,1,1,2,0,0,0},G_{1,0,1,1,0,1,1,2,1,0,0,0},
\nn\\
G_{1,0,1,1,0,1,1,2,2,0,0,0},G_{1,0,1,1,0,2,1,1,1,0,0,0}\}.
\nn
\end{eqnarray}
This can be achieved by choosing $G_{0,2,1,1,0,1,1,1,1,0,0,0} $ instead of $ G_{0,1,1,1,0,1,1,1,1,0,0,0}$.  
 
At level~8, one bad denominator appears,
$8 m^4-4 d m^4+24 m^2 s-5 d m^2 s-5 s^2+d s^2$.
We have to make changes in the two sectors with the following MIs:
\begin{eqnarray}
\{  
G_{0,1,1,1,1,1,1,1,1,0,0,0},G_{0,1,1,1,1,1,1,1,2,0,0,0},G_{0,1,1,1,1,1,1,2,1,0,0,0},G_{0,1,1,1,1,2,1,1,1,0,0,0},
\nn \\
 G_{1,0,1,1,1,1,1,1,1,0,0,0},G_{1,0,1,1,1,1,1,1,2,0,0,0},G_{1,0,1,1,1,1,2,1,1,0,0,0},G_{1,0,1,1,2,1,1,1,1,0,0,0}
 \}.
\nn
\end{eqnarray}
This can be achieved by choosing $G_{0,2,1,1,1,1,1,1,1,0,0,0}$ instead of $G_{0,1,1,1,1,1,1,1,1,0,0,0}$ 
and $G_{2,0,1,1,1,1,1,1,1,0,0,0}$ instead of $G_{1,0,1,1,1,1,1,1,1,0,0,0}$.

Finally, our code reveals no bad denominators at level~9.

\vspace{0.2 cm}
{\em Acknowledgments.}
We are grateful to J.~Davies, E.~Panzer, M.~Steinhauser and J.~Usovitsch for fruitful discussions
and various pieces of advice.
The work is carried out according to the research program of Moscow Center of Fundamental and Applied Mathematics.
 
\biboptions{longnamesfirst,sort&compress}
\bibliographystyle{elsarticle-num}
\bibliography{imi-v5}

\end{document}